\RequirePackage{fix-cm} 
\documentclass[journal=jacsat,manuscript=article,layout=twocolumn]{achemso}
\usepackage[fontsize=10pt]{fontsize}

\usepackage{amsmath}
\usepackage{amssymb}
\usepackage{subcaption}
\usepackage{bm}
\usepackage{xr}

\newcommand{\imgext}{png}
\newcommand{\al}[1]{\begin{align*}#1\end{align*}}
\newcommand*\aln[1]{\begin{align}#1\end{align}}

\author{Felix Post}
\affiliation[]{Max Planck Institute for Polymer Research, Ackermannweg 10, 55128 Mainz, Germany}
\email{felix.post@mpip-mainz.mpg.de}

\author{Jean-Philip Filling}
\affiliation[]{Institute of Computer Science, Johannes Gutenberg-University, 55128 Mainz, Germany}

\author{Toulik Maitra}
\affiliation[]{Max Planck Institute for Polymer Research, Ackermannweg 10, 55128 Mainz, Germany}
\alsoaffiliation[]{University of California, Davis, 1 Shields Ave, CA 95616, USA}

\author{Falk May}
\affiliation[]{Merck Electronics KGaA, Frankfurter Str. 250, 64293 Darmstadt, Germany}

\author{Michael Wand}
\affiliation[]{Institute of Computer Science, Johannes Gutenberg-University, 55128 Mainz, Germany}

\author{Denis Andrienko}
\affiliation[]{Max Planck Institute for Polymer Research, Ackermannweg 10, 55128 Mainz, Germany}
\email{denis.andrienko@mpip-mainz.mpg.de}

\title{Ab initio parametrization of distributed polarizable force fields}
\date{\today}

\begin{document}


\begin{abstract}
    Polarizable force fields offer superior transferability and accuracy compared to classical force fields, enabling access to electronic response properties such as refractive index and electronic density of states. Here, we demonstrate two key improvements that significantly enhance their accuracy: (1) assigning atomic polarizability to individual atoms rather than atom types, and (2) employing atomic polarizability tensors instead of scalar values. These modifications extend the applicability of polarizable force fields to cations, anions, and excited states, while also providing more accurate descriptions of neutral molecules.
    We propose a first-principles-based parameterization procedure for atomic polarizability tensors and scalars, validated on a set of small organic molecules with conjugated building blocks. To overcome the computational cost of ab initio calculations, we train a message-passing graph neural network to predict polarizability parameters, enabling efficient and scalable parameterization.
    Crucially, this approach imposes no additional computational cost during simulations and provides a clear diagnostic criterion for identifying cases where polarizable force field models fail to accurately describe molecular polarizability.
\end{abstract}

\thispagestyle{empty}

\section{Introduction}
Predicting material behavior at the atomic scale is a central goal in computational materials design. Classical molecular dynamics (CMD) offers an effective balance between computational efficiency and predictive accuracy for this purpose. CMD simulations propagate the equations of motion using a potential energy surface (PES), commonly referred to as a force field. The accuracy, complexity, and computational efficiency of the PES are inherently interconnected~\cite{muser_interatomic_2023}.
Simple PES models rely solely on pairwise interactions -- such as Lennard-Jones (LJ) and Coulomb potentials -- which approximate dispersion forces, exchange repulsion, and electrostatic interactions. However, these models often lack transferability, requiring frequent reparameterization of partial charges or LJ parameters for different system states or environments. To address these limitations, polarizable force fields have been developed to simultaneously enhance transferability and accuracy, particularly in systems where heterogeneous electrostatic environments play a critical role. These models are essential for accurately describing phenomena such as ion permeation through membranes~\cite{chen_molecular_2021}, energetic disorder and broadening of electronic-state distributions in molecular solids~\cite{reiser_analyzing_2021,kaklamanis_polar_2025}, and charge separation at interfaces with varying dielectric properties~\cite{ryno_polarization_2016,kaklamanis_polar_2025}. Polarizable force fields incorporate many-body interactions that are non-additive, making them computationally more demanding than traditional pairwise force fields. Consequently, it is essential to maximize model accuracy while maintaining its computational cost. Recent $\Delta$-machine-learning work has further shown that dipole-interaction models provide useful physically motivated baselines for molecular polarizability tensors, while also pointing to atom-specific polarizabilities as a route to better account for local environments~\cite{chaudhry_delta_2025}.

Because the present work builds directly on the structure of damped induced-dipole models, we introduce the relevant theoretical background alongside the motivation. One of the first polarizable force fields was proposed by Applequist~\cite{applequist_atom_1972}, which assigns an atomic polarizability tensor $\bm{\alpha}_p$ to each atom $p$ in a molecule. When the molecule is placed in an external electric field, each atom develops a local induced dipole ${ \bm \mu}_p$. This induced dipole arises from the external field $\bm{E}_p$ at the position of atom $p$, as well as the field generated by all other induced dipoles, given by $- \sum_{q\ne p}\bm{T}_{pq} \cdot \bm{\mu}_q $, where $\bm{T}_{pq}$ is the dipole field tensor at atom $p$ due to a dipole at atom $q$. The induced dipoles are determined by solving the following self-consistent equations
\aln{
\bm{\mu}_p = \bm{\alpha}_p\left( \bm{E}_p - \sum_{q\ne p}\bm{T}_{pq}\cdot \bm{\mu}_q \right).
\label{eq:mu_ind_p}
}

The structure of Eq.~\ref{eq:mu_ind_p} shows that the quality of the induced-dipole model is governed by two central components: the interaction tensor $\bm{T}_{pq}$ and the atomic polarizabilities $\bm{\alpha}_p$. Because the use of the bare dipole-dipole interaction tensor leads to the so-called polarization catastrophe, Thole~\cite{thole_molecular_1981} proposed scaling down large interaction energies at short distances by modifying the interaction tensors with distance-dependent screening functions that correspond to a Gaussian-smeared charge distribution. Many different functional forms thereof are applied and reported in the literature.~\cite{van_duijnen_molecular_1998,ren_polarizable_2011} Jensen et al.~\cite{jensen_polarizability_2002} further refined this approach, proposing a modified method in which the interatomic distances, $\bm{r}_{pq} = \bm{r}_{q} - \bm{r}_{p}$, entering the interaction tensor are rescaled. The dipole interaction tensor retains its original form,
\al{
\bm{T}_{pq} = \dfrac{\bm{1}}{{s}^3_{pq}}-\dfrac{3}{{s}^5_{pq}}(\bm{s}_{pq}\otimes\bm{s}_{pq}),
}
where $\bm{s}_{pq}=f(r_{pq})\cdot \bm{r}_{pq}/r_{pq}$ is the interatomic unit vector scaled by the distance-dependent function $f(r_{pq})$, and $\otimes$ denotes the outer product. In this work we adopt the following scaling function,
\al{
f(r_{pq}) = \sqrt{r_{pq}^2+\dfrac{\pi}{4}\cdot \dfrac{\Phi_p+\Phi_q}{\Phi_p\cdot \Phi_q}},
}
which corresponds to the IM-SQRT model introduced by Jensen~\cite{jensen_polarizability_2002}. Note that, in contrast to Thole's approach, where screening functions depend only on atomic polarizabilities $\alpha_p$ and a universal damping parameter, atom-specific damping parameters, $\Phi_p$, improve the transferability across different chemical environments and enhance the robustness of the model.

As proposed by Thole and Jensen, polarizable force fields typically assign polarizabilities by atom type, using identical scalar values for chemically similar sites, and fit these parameters using a dataset of molecules with experimentally measured molecular polarizabilities. This approach reduces the number of parameters but also limits the model to neutral states of small, non-conjugated molecules.~\cite{chaudhry_delta_2025} In addition, scalar atom-type parameters cannot fully resolve local chemical environments, anisotropic electronic response, or state-dependent changes in molecular polarizability.

In the present work, we extend this framework: each atom is assigned a unique polarizability value (scalar or tensor), derived from first-principles calculations. This modification provides finer resolution, resulting in improved accuracy and transferability in modeling electrostatic response. It enables more accurate treatment of neutral molecules and extends the framework to molecular conformers, cations, anions, and electronically excited states -- all within a unified theoretical approach. Additionally, we explain why distributed polarizability models are (surprisingly) accurate, despite neglecting changes in atomic partial charges in response to external fields.

The paper is organized as follows. We first describe the fitting procedure and explain how charge flow can be incorporated into induced dipoles. We then parameterize atomic polarizabilities for a diverse set of compounds with varying molecular weights, including anionic, cationic, and triplet and singlet excited states. The model's accuracy is demonstrated by benchmarking against molecular polarizability tensors. We further assess the conformational transferability of the fitted parameters by applying atomic polarizabilities fitted at one TPBi reference geometry to representative conformers of the same molecule. The resulting molecular polarizability tensors are then used to compute refractive indices for comparison with experimental data. Finally, we train an equivariant message-passing graph neural network to predict atomic polarizabilities, providing an efficient alternative to computationally demanding first-principles calculations.

\section{Results and Discussion}

\subsection{The fitting and parameterization algorithm}
The proposed algorithm can be summarized as follows: For a given molecule, we evaluate its electron density in a uniform external field applied in the $x$, $y$, and $z$ directions, using density functional theory, details in the Methods section. When using a uniform external field, the atomic index of the field in Eq.~\ref{eq:mu_ind_p} can be neglected since the external contribution is the same at each atom ($\bm{E}_p=\bm{E}$). The electron density is then partitioned onto atomic multipoles using the distributed multiple analysis (DMA). We use density derived electrostatic and chemical electron density partitioning (DDEC6)~\cite{manz_2016_1,manz_2016_2,manz_2016_4} but other partitioning schemes, e.g., GDMA~\cite{stone_gdma_2005}, can also be used. Details are provided in the methods section. 

DMA provides a set of induced dipoles per atom, $\bm{\mu}^{\text{qm}}_p = \bm{\mu}_p(\bm{E}) - \bm{\mu}_p(\bm{E}=0)$.  At first glance, it might seem sufficient to find a set of atomic polarizabilities that closely reproduce $\bm{\mu}^{\text{ref}}_p$. However, this approach is inherently flawed and will fail to accurately reproduce either the molecular polarizability or the total induced dipole of the molecule. A closer examination of the multipole expansion reveals that the external field not only induces atomic dipoles but also alters the partial charges. This charge redistribution contributes an additional term to the induced molecular dipole, which we term the charge flow dipole, $\bm{\mu}^\text{cf}$.

The Thole/Jensen distributed polarizability models account for this charge flow \textit{effectively} by distributing it among atomic induced dipoles -- the fit to molecular polarizabilities automatically captures this effect. However, this approach is not feasible in our case, as we have significantly more polarizability parameters per molecule to fit, necessitating the explicit use of both induced atomic dipoles and charges. To address this, we adopt a workaround inspired by the bond charge approach~\cite{heid_evaluating_2018}, converting the charge response $q^{\text{qm}}_p = q_p(\bm{E}) - q_p(\bm{E}=0)$ into atomic charge flow dipoles, $\bm{\mu}^{\text{cf}}_p$, as detailed in the Supporting Information.


The final cost function $\Phi(\{\bm{\alpha}_p\})$, minimized using the bound-constrained limited-memory BFGS algorithm (L-BFGS-B), incorporates both contributions,
\aln{
\Phi = \sum_{p}\sum_{\bm{E}\in \{E_x, E_y, E_z\}}
\left|
    \bm{\mu}_p^{\text{qm}} + \bm{\mu}_p^{\text{cf}} - \bm{\mu}_p(\bm{\alpha}_p) 
\right|^2 ,
\label{fit-function_final}
}
where the summation runs over all atoms $p$ and over the three Cartesian field directions, $E_x$, $E_y$, and $E_z$ of $\bm{E}$ seperately.

\subsection{Molecular dataset}

\begin{figure*}
    \includegraphics[width=0.8\textwidth]{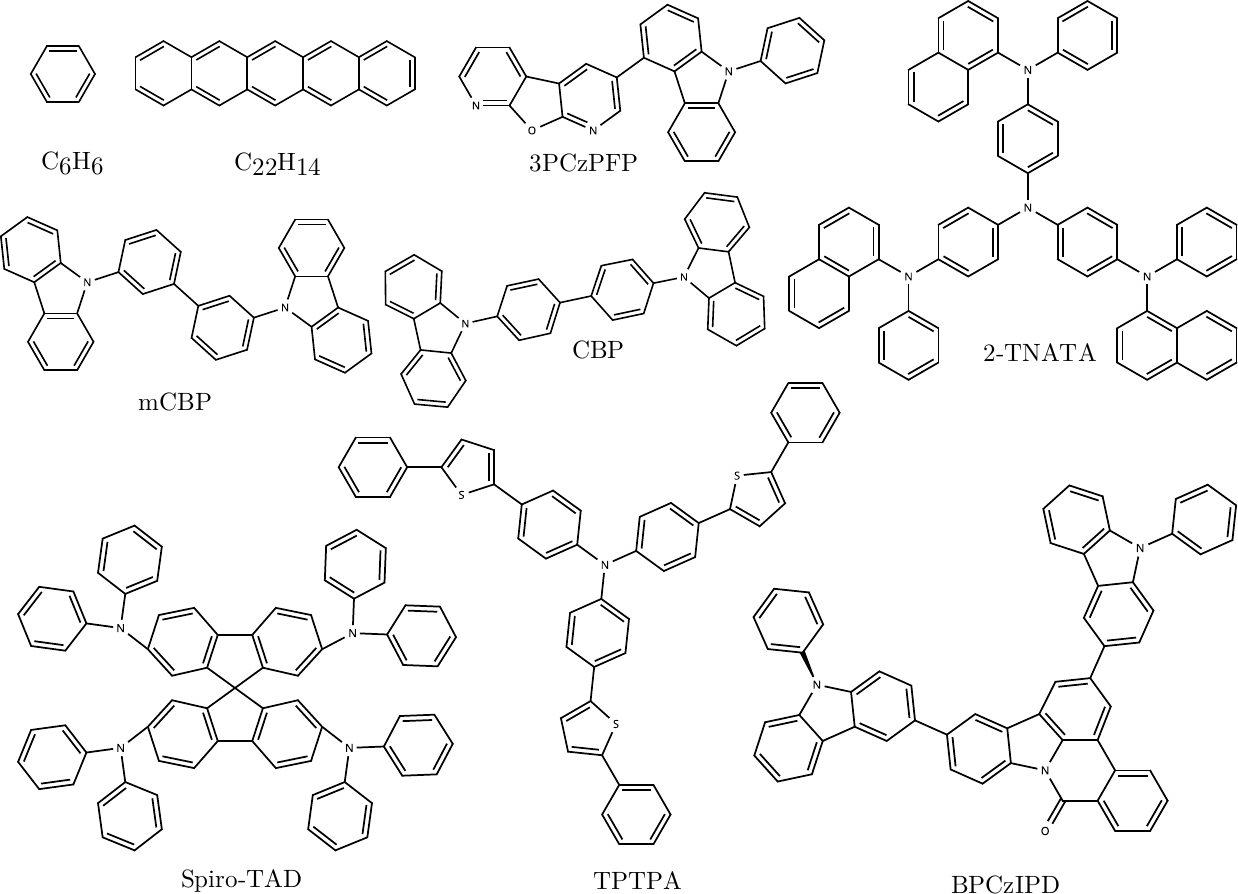}
    \caption{A representative subset of typical molecules from the molecular dataset is shown.}
    \label{fig:structures}
\end{figure*}

The accuracy of the proposed parametrization was benchmarked on a diverse set of 76 organic molecules, with 80\% comprising host or emitter materials commonly used in organic light-emitting diodes (OLEDs). To further enhance the dataset's diversity, the remaining 20\% includes diatomic molecules and standard organic fragments such as benzene and triazine. A representative subset -- illustrated in Figure~\ref{fig:structures} -- was selected to span variations in molecular size (number of atoms), dipole moment strength, average polarizability tensor, and its anisotropy. The complete dataset is available for download from the online repository, as detailed in the Data Availability Section.


\subsection{Neutral molecules}

Thole-Jensen models were parametrized and benchmarked on neutral states of molecules, with scalar atomic polarizabilities. To assess the added value of having a polarizability per atom and not atom type, we first minimized the cost function for every neutral molecule in the dataset and calculated the molecular polarisability $\bm \alpha_{\text{mol}}$ from the total induced dipole by inverting the matrix equation 
$\sum_p \bm{\mu}_p = \bm{\alpha}_{\text{mol}} \cdot \bm{E} $.

\begin{figure*}
    \includegraphics[width=0.8\textwidth]{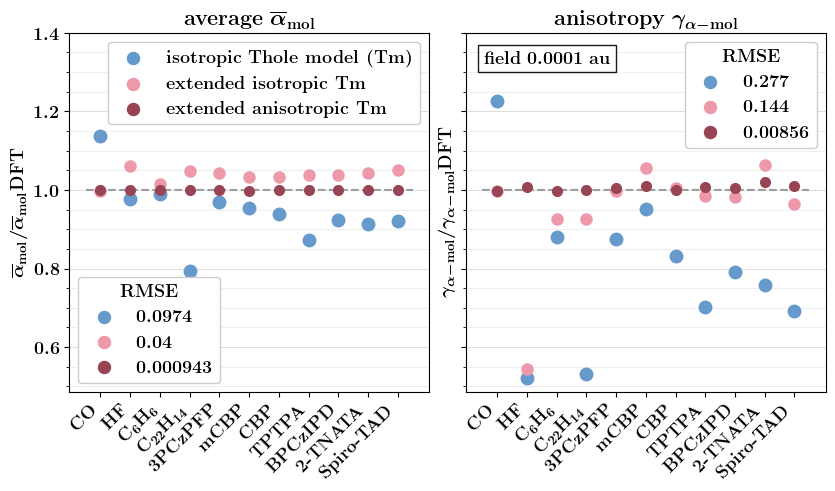}
    \caption{Representative subset of neutral molecules orderd and selected according their sizes. The diatom CO being the smallest whereas Spiro-TAD contains 129 atoms. The corresponding figures of the same, but singly charged, molecules are included in the SI (Section~\ref{sec:representative_anion_cation}).}
    \label{fig:subset_mean_aniso_pp0}
\end{figure*}

The main quantities compared throughout this work are the simulated and model-predicted average molecular polarizability, $\bar{\alpha}_{\text{mol}}$, and the rotationally invariant anisotropy, $\gamma_{\alpha\text{-mol}}$, which is the Frobenius norm of the traceless part of the polarizability tensor $\bm \alpha_{\text{mol}}$ in its standard Placzek/Raman form~\cite{placzek_intensitaet_1931,long_raman_2002,sunaga_vibrational_2026}.
\begin{equation}
    \bar{\alpha} = \dfrac{\operatorname{Tr}{\left(\bm{\alpha}\right)}}{3},\qquad \gamma = \sqrt{\dfrac{3}{2}}\, \big\lVert \bm{\alpha} - \bar{\alpha}I_3 \big\rVert_{F} \label{eq:avg_and_aniso_pol}
\end{equation}
In Eqs.~\ref{eq:avg_and_aniso_pol}, the subscripts $\text{mol}$ and $\alpha\text{-mol}$ have been omitted for brevity and $I_3$ is the $3\times 3$ identity matrix. These describing quantities according to the molecular polarisability tensor are plotted in Figure \ref{fig:subset_mean_aniso_pp0} using the atom-type specific atomic polarisabilities taken from Jensen \cite{jensen_polarizability_2002} in blue and the fitted values according to our new approach in light and dark red. 

In the case of fitted scalar polarisabilities (light red in Figure \ref{fig:subset_mean_aniso_pp0}) the new atom specific model shows perfect matching compared to DFT-derived values for medium and bigger sized molecules, whereas for small diatomic molecules a bigger deviation is present. For the small molecules, the electronic response of the charge distribution can not be fully described by scalar atomic polarizabilities. This is due to the small amount of parameters which can be easily expanded by using tensorial atomic polarisabilities (dark red in Figure \ref{fig:subset_mean_aniso_pp0}). 

\begin{figure*}
    \includegraphics[width=0.8\textwidth]{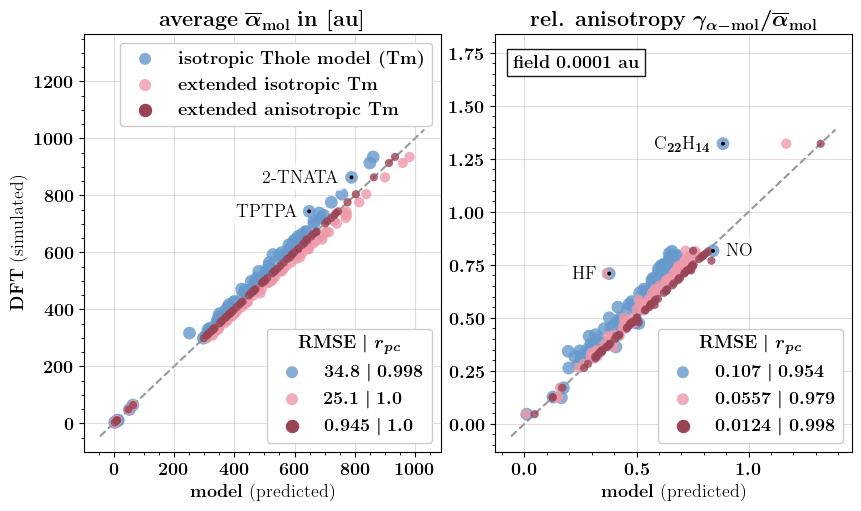}
    \caption{Benchmark of Thole-model molecular polarizabilities for 76 \textbf{neutral} molecules against DFT reference data. Left: parity plot of the isotropic mean polarizability $\bar{\alpha}_{\mathrm{mol}}$ (a.u.). Right: parity plot of the relative anisotropy $\gamma_{\alpha\text{-mol}}/\bar{\alpha}_{\mathrm{mol}}$ obtained from the molecular polarizability tensor. Predictions are shown for the atom-type isotropic Thole model using Jensen parameters (blue), the extended isotropic variant (light red), and the extended anisotropic variant (dark red). Insets summarize the prediction errors and correlations according to eq.~\ref{eq:def_errors} (SI).}
     \label{fig:neutral}
\end{figure*}

Besides pentacene (C$_{22}$H$_{14}$) and TPTPA the atom-type specific model (blue in Figure \ref{fig:subset_mean_aniso_pp0}) shows a systematic underestimation of the average polarisabilieties which can be related to a different level of theory used in the parametrization data of the atom-specific model compared to ours. This is also evident for the entire data set shown in Figure \ref{fig:neutral}, where we use a relative anisotropy $\bar{\alpha}_{\text{mol}}/\gamma_{\alpha\text{-mol}}$ for comparison.

A uniform rescaling can reduce this bias in the mean response, yet it does not address the spread in anisotropy, indicating that the missing information is not a simple prefactor but rather molecule-dependent directional response. Allowing atom-resolved polarizabilities markedly improves both metrics and yields near-parity behavior across most of the chemical space. The remaining discrepancies are most apparent for very small systems, where constraining the sites to scalar polarizabilities limits expressivity. Our new atom-specific and tensorial model overcomes this issue and describes the magnitude and direction dependence of the polarisability on a much presicer level.

\subsection{Charged molecules}

\begin{figure*}
    \includegraphics[width=0.8\textwidth]{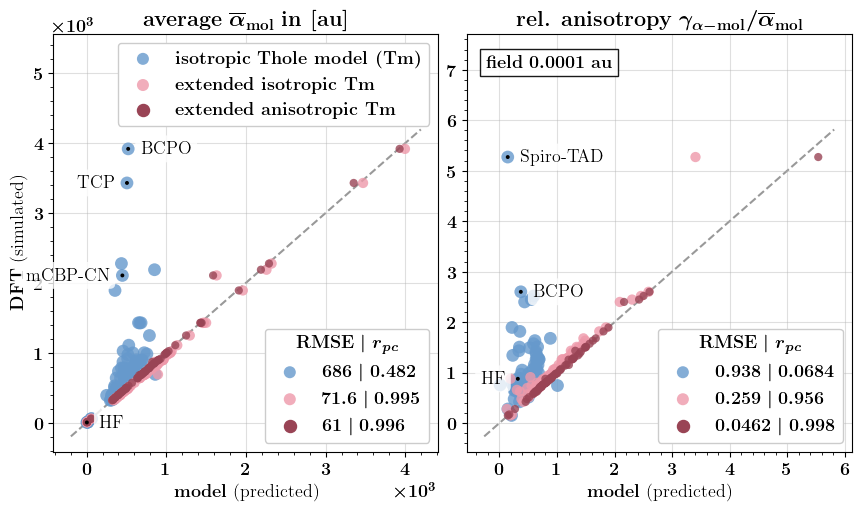}
    \caption{[$\mathbf{\lambda=0}$] Benchmark of Thole-model molecular polarizabilities for 75 \textbf{singly charged cations} (mCBP excluded) against DFT reference data. Layout and symbols as in Figure \ref{fig:neutral}.}
    \label{fig:cations}
\end{figure*}

Thole-- and Jensen--type parameterizations are typically derived for neutral molecules and their direct transferability to charged electronic states is not guaranteed. To check the performance for singly charged cations, we repeat the per-molecule optimization protocol and extract the molecular polarizability tensor using the same field strength ($10^{-4}$~au) as for the neutral benchmark and compare the mean polarizability in addition to the relative anisotropy against the DFT reference (Figure~\ref{fig:cations}).

For the cations, the atom-type isotropic Thole model shows markedly reduced predictive power, with substantial scatter in $\bar{\alpha}_{\mathrm{mol}}$ (RMSE $=686$, $r_{pc}=0.482$) and essentially no correlation for the anisotropy measure (RMSE $=0.938$, $r_{pc}=0.0684$). In contrast, both extended variants restore near-parity behavior: the extended isotropic model achieves RMSE $=71.6$ ($r_{pc}=0.995$) for $\bar{\alpha}_{\mathrm{mol}}$ and RMSE $=0.259$ ($r_{pc}=0.956$) for the anisotropy. Introducing anisotropic extensions further improves the directional response, reducing the anisotropy error to RMSE $=0.0462$ with $r_{pc}=0.998$, while maintaining excellent agreement for $\bar{\alpha}_{\mathrm{mol}}$ (RMSE $=61$, $r_{pc}=0.996$). Remaining outliers (e.g., Spiro-TAD and BCPO in the anisotropy panel) indicate that a small number of ions still exhibit response characteristics that are challenging to represent with scalar site polarizabilities alone.

\begin{figure*}
    \includegraphics[width=0.8\textwidth]{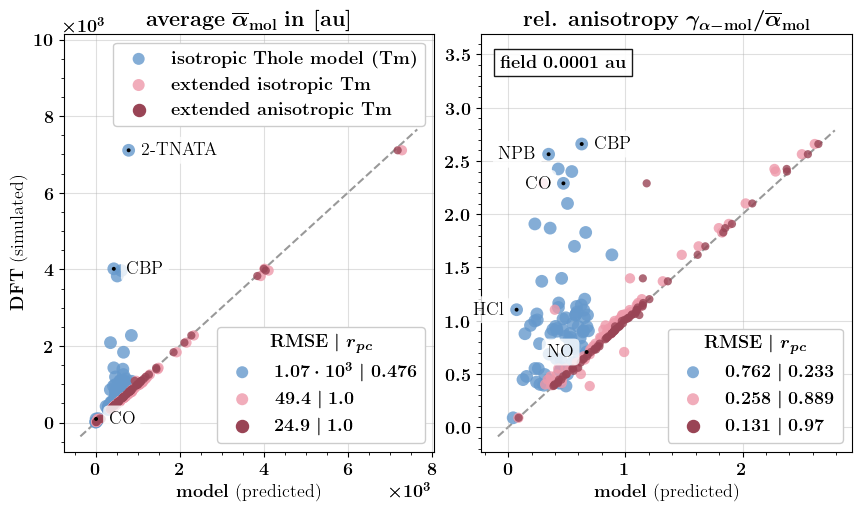}
    \caption{[$\mathbf{\lambda=0}$] Benchmark of Thole-model molecular polarizabilities for 76 \textbf{singly charged anions} against DFT reference data. Layout and symbols as in Figure \ref{fig:neutral}.}
    \label{fig:anions}
\end{figure*}

An analogous benchmark was carried out for singly charged anions (Figure~\ref{fig:anions}). Compared to the neutral molecules, the atom-type isotropic Thole model deteriorates even more strongly for anions, yielding RMSE $=1.07\times 10^{3}$ ($r_{pc}=0.476$) for $\bar{\alpha}_{\mathrm{mol}}$ and RMSE $=0.762$ ($r_{pc}=0.233$) for the relative anisotropy. Several prominent deviations (e.g., 2-TNATA and CBP in $\bar{\alpha}_{\mathrm{mol}}$, as well as NPB/CBP in the anisotropy panel) underline that a fixed atom-type assignment does not adequately capture the environment dependence of the response in the anionic state.

The extended models substantially improve the agreement with DFT. For $\bar{\alpha}_{\mathrm{mol}}$, the extended isotropic and extended anisotropic variants reach RMSE values of $49.4$ and $24.9$, respectively, with essentially perfect correlation ($r_{pc}=1.0$ in both cases). For the anisotropy, the improvement is similarly pronounced: RMSE decreases from $0.762$ (atom-type) to $0.258$ (extended isotropic, $r_{pc}=0.889$) and further to $0.131$ for the extended anisotropic model ($r_{pc}=0.97$). Overall, allowing atom-specific degrees of freedom is therefore critical to recover both magnitude and directional trends for anions.

Across all charge states, the atom-type isotropic parametrization shows the weakest transferability, with performance degrading substantially from neutrals to ions---most dramatically for the anisotropy (neutrals: RMSE $\approx 0.107$, $r_{pc}\approx 0.954$; cations: RMSE $=0.938$, $r_{pc}=0.0684$; anions: RMSE $=0.762$, $r_{pc}=0.233$). The extended models largely restore near-parity behavior for $\bar{\alpha}_{\mathrm{mol}}$ in both cations and anions (RMSE $\sim 25$--$72$, $r_{pc}\gtrsim 0.995$), indicating that the magnitude of the response can be captured once sufficient molecule-specific flexibility is introduced. However, reproducing the directional dependence remains more demanding for charged systems: even the best anisotropy agreement is obtained for cations (extended anisotropic RMSE $=0.0462$, $r_{pc}=0.998$), while anions remain less accurate (RMSE $=0.131$, $r_{pc}=0.97$) and neutrals are overall easiest (RMSE $\approx 0.0124$, $r_{pc}\approx 0.998$). Charged states appear more sensitive to the local chemical environment and its coupling to long-range induction, which is difficult to capture when sites are restricted to scalar polarizabilities. Introducing molecule-specific (atom-resolved) polarizabilities improves both the overall magnitude and the directional response for cations and anions. In our data set, anions remain slightly more challenging in terms of anisotropy than cations, indicating a stronger environment dependence.

\subsection{Excited States}

Compared to ground states, excited-state polarizabilities remain far less explored territory.\cite{grozema_excited_2001, heid_evaluating_2018, zhao_excited-state_2023} While ground-state polarizabilities of organic molecules are routinely computed and measured, systematic studies of S$_1$ and T$_1$ state polarizabilities are scarce, and most existing work focuses on molecular-level properties rather than atomic decomposition. 

\begin{figure*}
    \includegraphics[width=0.8\textwidth]{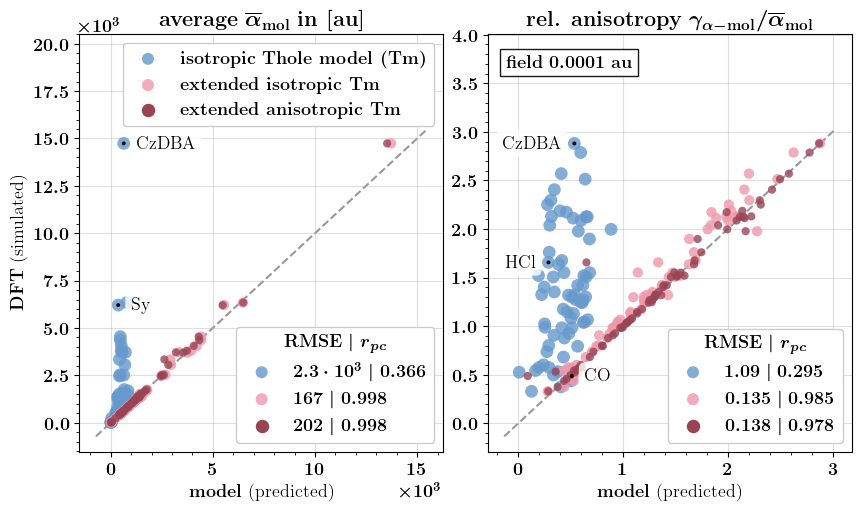}
   \caption{Comparison of predicted versus DFT-calculated molecular polarizabilities for the first singlet excited state (S$_1$). \textbf{Left:} average polarizability $\bar{\alpha}_{\text{mol}}$. \textbf{Right:} relative anisotropy $\gamma/\bar{\alpha}_{\text{mol}}$. Blue points represent the atom-type model from Ref.~\citenum{jensen_polarizability_2002}; red points represent the atom-specific model developed in this work. The dashed line indicates perfect agreement.}
    \label{fig:singlet}
\end{figure*}

\begin{figure*}
    \includegraphics[width=0.8\textwidth]{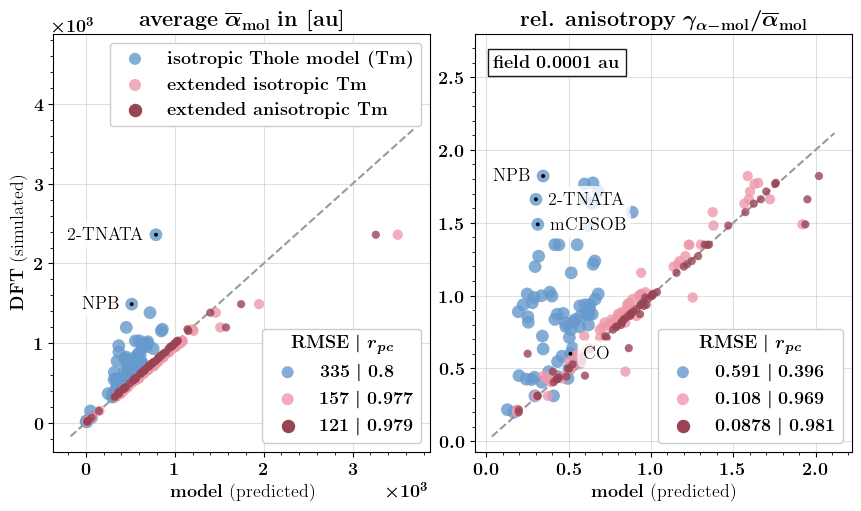}
    \caption{Comparison of predicted versus DFT-calculated molecular polarizabilities for the first triplet excited state (T$_1$). Layout and symbols as in Figure~\ref{fig:singlet}. Note the differences in anisotropy distribution compared to the S$_1$ state, reflecting the distinct electronic structure of triplet excited states.}
    \label{fig:triplet}
\end{figure*}

Excited-state polarizabilities from TDDFT are highly sensitive to the choice of exchange–correlation functional and may become unreliable in the presence of charge-transfer errors or triplet instabilities. Well-known TDDFT failures for long-range charge-transfer excitations and instability-related triplet states can lead to qualitatively incorrect excitation energies and consequently unphysically large response properties.\cite{dreuw_longrange_2003,peach_influence_2011} Consequently, TBA, TPTPA, TBPT, and TCP were omitted from the excited-state analysis because at the choosen level of theory the computed mean molecular polarizabilities ($\bar{\alpha}_{\text{mol}}$) exceeded $100\,000\,\mathrm{au}$, suggesting numerically unstable or unphysical response behavior. Figures~\ref{fig:singlet} and \ref{fig:triplet} compare the predicted molecular polarizabilities for S$_1$ and T$_1$ states, respectively, against DFT reference values.

Two molecules exhibit substantial deviations from the triplet reference values, TCTA in terms of average polarizability and mCBP in terms of relative anisotropy (TCTA $\Delta \bar{\alpha}_{\text{mol}}>4300\,\mathrm{au}$, mCBP $\Delta(\bar{\alpha}_{\text{mol}}/\gamma_{\alpha\text{-mol}})>3.3$). These molecules were excluded from the correlation shown in Figure \ref{fig:triplet} for clarity but are included in the full dataset shown in Figure \ref{fig_si:triplet_all}. Including all molecules results in an RMSE of the ext. anisotropic model of $655\,\mathrm{au}$ compared to $120\,\mathrm{au}$ for the truncated dataset. The noticeable deviations appear to originate from delocalized orbitals, indicating a specific limitation of the present model.

The atom-type model (blue) exhibits systematic and severe underestimation of both average polarizability ($\bar{\alpha}_{\text{mol}}$) and relative anisotropy ($\bar{\alpha}_{\text{mol}}/\gamma_{\alpha\text{-mol}}$) across the majority of molecules in the dataset. This failure is more pronounced than observed for ground states (Figure~\ref{fig:neutral}) and can be attributed to the fundamentally different electronic structure of excited states. In these states, electron density becomes more diffuse, charge distributions reorganize significantly, and in many molecules, particularly those with donor-acceptor units, substantial charge-transfer contributions emerge. Under these conditions, assigning identical polarizabilities to chemically similar atoms becomes untenable, as the local electronic environment at each atom differs markedly from the ground state.

In contrast, our atom-specific model (red) accurately reproduces both the magnitude and directional dependence of excited state polarizabilities. For the average polarizability, we observe excellent agreement with DFT across the entire dataset, with Pearson correlations of $r_{pc} > 0.977$ for both S$_1$ and T$_1$ states. The model also correctly captures the relative anisotropy characteristic of excited states. This enhanced anisotropy reflects the directional nature of electronic excitations, particularly in conjugated systems where excitation promotes electron density along specific molecular axes.

Comparing S$_1$ and T$_1$ states reveals physically meaningful differences between singlet and triplet manifolds. While both show enhanced polarizabilities relative to the ground state, a consequence of more diffuse electron densities, the distributions and magnitudes of the relative anisotropies differ. For the S$_1$ state, molecules with strong charge-transfer character exhibit the largest anisotropies, reflecting the directional redistribution of electron density upon excitation. In contrast, T$_1$ states, which often have more localized triplet densities due to exchange stabilization, show a different anisotropy distribution. This aligns with earlier findings by Grozema et al., who reported that triplet excitons in conjugated oligomers have smaller polarizabilities than singlet excitons, consistent with their more compact spatial extent.\cite{grozema_excited_2001}  The observation that $\bar{\alpha}_{\text{mol}}/\gamma_{\alpha\text{-mol}}(\text{S}_1) \neq \bar{\alpha}_{\text{mol}}/\gamma_{\alpha\text{-mol}}(\text{T}_1)$, indicates that the directional polarization response is sensitive to spin multiplicity, an effect that would be completely missed by ground-state-only parameterizations or by atom-type models that cannot adapt to different electronic configurations.

The importance of the charge flow contribution $\bm{\mu}^{\text{cf}}$ becomes particularly evident in excited states. As discussed in the parametrization algorithm, the external field not only induces atomic dipoles $\bm{\mu}_p^{\text{qm}}$ but also redistributes partial charges $\Delta q_p$, which contributes an additional term to the total molecular dipole. In excited states, this charge redistribution is amplified relative to the ground state due to the altered electronic structure and reduced HOMO-LUMO gap.\cite{ayers_physical_2006} Neglecting the charge flow term in Eq.~\ref{fit-function_final} results in systematic errors in predicted molecular polarizabilities for molecules with pronounced charge-transfer character, such as donor-acceptor chromophores commonly used in OLEDs.

In summary, these results demonstrate that atom-specific polarizabilities are not merely an incremental improvement but a necessity for excited state modeling. The ability to accurately predict excited state polarizabilities enables calculation of optical properties in excited state populations, modeling of excited state energetics in heterogeneous electrostatic environments, and prediction of electrostatic contributions to exciton--charge and exciton--exciton interactions in organic semiconductors.



\subsection{Transferability across molecular conformers}

After assessing the accuracy of the fitted atomic polarizabilities at the reference geometry, we next investigate how transferable these parameters are across different conformers of the same molecule. This test is important because molecular simulations sample a distribution of geometries, whereas a direct first-principles refit of the atomic polarizabilities for every molecular conformation would not be practical. Ideally, atomic polarizabilities fitted at a representative reference geometry should therefore remain applicable to nearby conformations, while the conformational dependence of the molecular polarizability tensor is captured through the geometry-dependent dipole-dipole interaction tensor.

As a representative test case, we selected TPBi, a well-known organic molecule in the field of organic optoelectronic host materials. TPBi is particularly suitable for this analysis because its extended aromatic structure and conformational flexibility make the molecular polarizability tensor sensitive to changes in molecular geometry, while the molecule remains chemically well defined and experimentally relevant. Initial conformer ensembles were generated from the corresponding XYZ structure using CREST~\cite{pracht_crest_2020} with the standard conformational search workflow at the GFN2-xTB level~\cite{bannwarth_gfn2_2019}. From this ensemble, five representative conformers were selected. The resulting conformer geometries were used without further geometry optimization for single-point DFT calculations in Gaussian 16 at the M06-2X/6-311+G(d,p) level. Relative SCF electronic energies, $\Delta E_{\text{SCF}}$, were calculated from the resulting single-point energies.

\begin{figure*}
    \includegraphics[width=.9\textwidth]{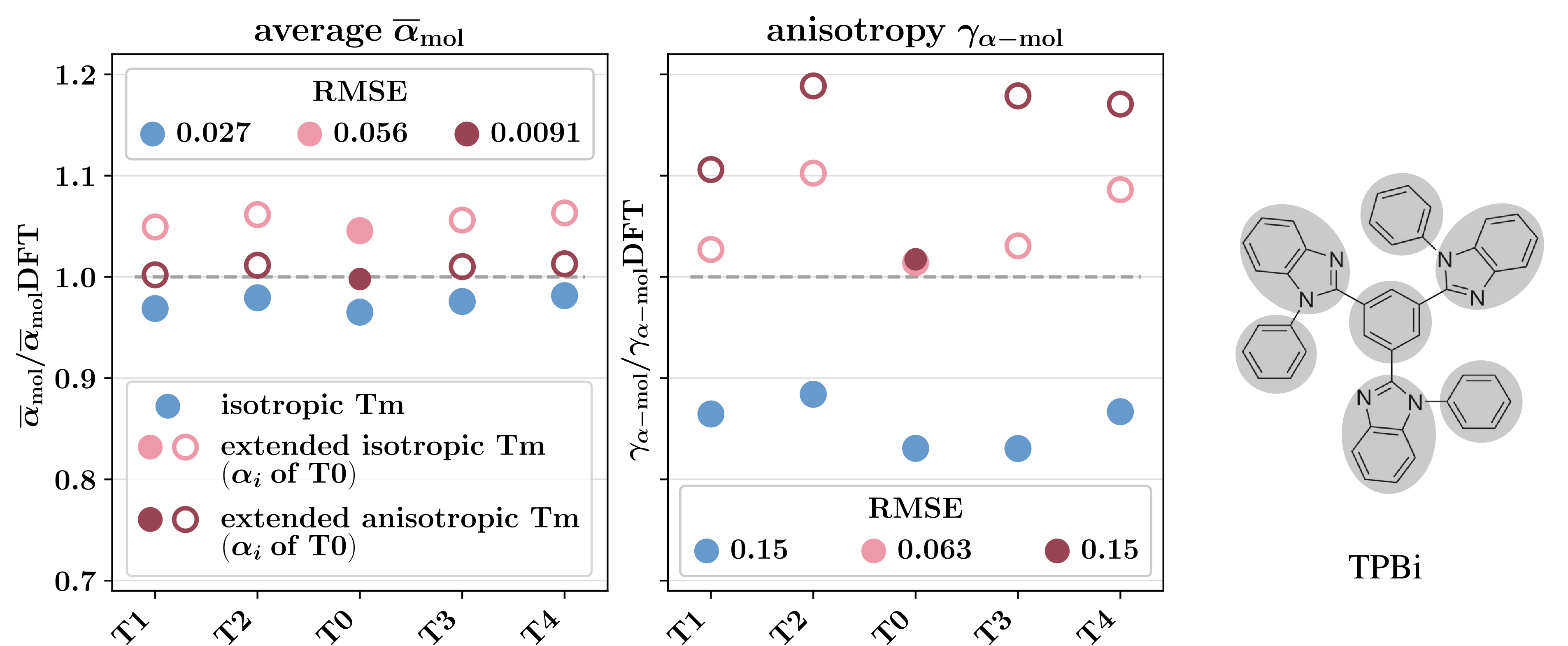}
    \includegraphics[width=\textwidth]{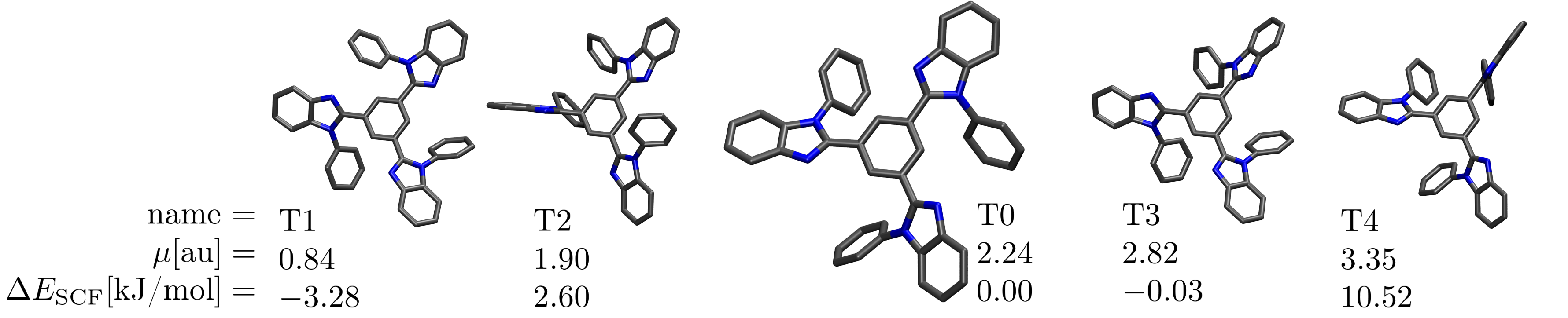}
    \caption{
Conformational transferability of fitted atomic polarizabilities for TPBi.
\textbf{Upper panel:} Relative deviations of the mean molecular polarizability and molecular polarizability anisotropy for the five conformers within the original isotropic Thole model (blue), the atom-resolved scalar fit (light red), and the atom-resolved tensorial fit (dark red). The local fragment frames, used to rotate the atomic polarizability tensors between conformers in the tensorial approach, are indicated as grey shaded areas in structure of TPBi on the right side.
\textbf{Lower panel:} Five representative TPBi conformers selected from the CREST conformer ensemble, together with their total electric dipole moments $\mu$ and relative SCF energies $\Delta E_{\mathrm{SCF}}$.
}
    \label{fig:conformer_transferability}
\end{figure*}

For TPBi, the atomic polarizabilities were fitted at one reference geometry (T0) and subsequently applied to the four representative conformers (T1-T4) without further optimization. For each conformer, the molecular polarizability tensor was recomputed using the conformer-specific atomic coordinates and the same damped induced-dipole model as used in the fitting procedure for T0. In the case of scalar atomic polarizabilities, i.e., the extended isotropic model, the polarizability value assigned to a given atom is invariant under molecular rotations and was therefore transferred directly to all conformers. For tensorial atomic polarizabilities, however, the orientation of each atomic tensor has to follow the corresponding molecular fragment. To this end, fragment-local coordinate frames were constructed for the reference geometry and for each conformer, as illustrated in the upper right panel of Fig.~\ref{fig:conformer_transferability}. The atomic polarizability tensor fitted in T0 was first expressed in its local fragment frame and then rotated into the corresponding local frame of each conformer. Equivalently, the transferred tensor for atom $p$ in conformer $k$ can be written as
\al{
\bm{\alpha}_{p}^{(k)} = \bm{\mathcal{R}}_{p}^{(k)} \bm{\alpha}_{p}^{(0)} \left(\bm{\mathcal{R}}_{p}^{(k)}\right)^{\mathrm{T}},
\label{eq:alpha_rotation}
}
where $\bm{\alpha}_{p}^{(0)}$ is the tensor fitted for the reference geometry T0 and $\bm{\mathcal{R}}_{p}^{(k)}$ maps the local frame of atom $p$ in T0 onto the corresponding local frame in conformer $k$. The rotation matrices, $\bm{\mathcal{R}}_{p}^{(k)}$, are obtained using the Kabsch algorithm~\cite{kabsch_discussion_1978}, which minimizes the root mean squared deviation between two paired sets of points. The resulting molecular polarizability tensors were then compared to the corresponding first-principles reference tensors calculated for the same conformers.

Figure~\ref{fig:conformer_transferability} shows the resulting comparison between the predicted and reference molecular polarizability descriptors for the five TPBi conformers. The left panel reports the relative deviation of the mean molecular polarizability, $\bar{\alpha}_{\mathrm{mol}}/\bar{\alpha}_{\mathrm{mol}}^{\mathrm{DFT}}$, while the right panel shows the corresponding deviation of the molecular polarizability anisotropy, $\gamma_{\alpha-\mathrm{mol}}/\gamma_{\alpha-\mathrm{mol}}^{\mathrm{DFT}}$. For the mean polarizability, the transferred atom-resolved tensorial polarizabilities give the most accurate results, with an RMSE of 0.0091. The original isotropic Thole model slightly underestimates $\bar{\alpha}_{\mathrm{mol}}$ for all conformers, whereas the extended isotropic fit systematically overestimates it. Thus, for the isotropic part of the molecular response, the tensorial atomic polarizability model provides the most transferable description across the considered TPBi conformers.

The behavior is different for the molecular polarizability anisotropy. Here, the original isotropic Thole model systematically underestimates the DFT anisotropy, while the transferred tensorial polarizabilities tend to overestimate it for most conformers. The extended isotropic fit gives the best overall agreement for $\gamma_{\alpha-\mathrm{mol}}$, with an RMSE of 0.063, compared with 0.15 for both the original isotropic Thole model and the transferred tensorial model. This indicates that a more accurate description of the mean molecular polarizability does not necessarily improve the conformational transferability of the molecular anisotropy, which is more sensitive to the relative orientation of polarizable molecular fragments.

These results clarify the role and limitations of fitted atomic polarizabilities in the present model. Atomic polarizabilities fitted at a single reference geometry should not be interpreted as geometry-independent universal constants, but as an efficient atom-resolved representation of the electronic response around the chosen reference structure. For TPBi, this representation is highly transferable for the mean molecular polarizability, especially when tensorial atomic polarizabilities are used. In contrast, the molecular anisotropy is more sensitive to conformational changes, and the scalar atom-resolved fit gives the most balanced description for the selected conformer set. The conformer transferability test therefore provides a practical diagnostic for identifying which molecular response properties can be described by a single set of fitted atomic polarizabilities and which may require conformation-dependent parameters or machine-learned polarizabilities.


\subsection{Refractive index}
Having assessed the transferability of the fitted atomic polarizabilities at the molecular level, we next check the predictive accuracy of the proposed multiscale polarization model by comparing simulated and experimental refractive-index anisotropies in vapor-deposited thin films. Because $\Delta n$ depends on both the orientational distribution in the deposited film and the anisotropic molecular polarizability, it provides a sensitive experimental benchmark for the present multiscale polarization model.

The simulated thin films were generated with the coarse-grained vapor-deposition and backmapping workflow introduced in our previous paper~\cite{scherer_predicting_2024}, yielding atomistically resolved morphologies for the subsequent $\Delta n = n_{\mathrm{eo}} - n_{\mathrm{o}}$ analysis.

We compare simulated $\Delta n$ values with ellipsometrically measured refractive-index anisotropies at 620~nm, where $n_{\mathrm{o}}$ denotes the ordinary refractive index in the $xy$-plane and $n_{\mathrm{eo}}$ the extraordinary refractive index along the $z$-axis. Thus, $\Delta n$ directly reflects the degree of optical anisotropy, with positive values indicating, on average, a more upright molecular orientation and negative values corresponding to predominantly flat-lying morphologies.

The refractive indices were subsequently evaluated using a Thole/Jensen-type polarizable dipole-interaction model. In contrast to our previous implementation, where atom-type specific atomic polarizabilities were adopted from the original IM-SQRT parameterization, the present $\Delta n$ calculations use atomic polarizabilities obtained from our induced-dipole fit to our DFT reference data. This substitution preserves the original functional form of the polarization model while making the optical response consistent with the level of theory used throughout this work.

\begin{figure}
    \centering
    \includegraphics[width=\columnwidth]{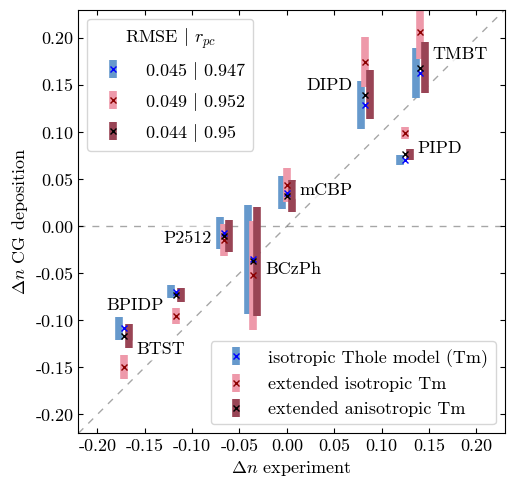}
    \caption{Comparison of experimental and simulated refractive-index anisotropies, $\Delta n = n_{\mathrm{eo}} - n_{\mathrm{o}}$, for vapor-deposited thin films. Results are shown for the original isotropic Thole model (Tm, blue), the atom-resolved isotropic fit (pink), and the atom-resolved anisotropic fit (dark red). Crosses denote mean values from five independently deposited films, and the vertical bars indicate the corresponding statistical uncertainty.}
    \label{fig:delta_n}
\end{figure}

Figure~\ref{fig:delta_n} compares simulated and experimental refractive-index anisotropies, $\Delta n = n_{\mathrm{eo}} - n_{\mathrm{o}}$, for the vapor-deposited thin films. The original isotropic Thole model, already validated in our previous work, reproduces the experimental birefringence well. The present atom-resolved isotropic and anisotropic fits therefore probe whether a more faithful molecular polarizability description improves the macroscopic optical response. All three models reproduce the correct sign of $\Delta n$ for all compounds and yield similarly strong agreement with experiment. While the extended isotropic fit gives the highest correlation coefficient ($r_{pc}=0.952$), the extended anisotropic fit yields the lowest RMSE (0.044), only slightly improving over the original isotropic model (0.045). Thus, the more accurate reproduction of the single-molecule reference data translates into only a modest gain at the thin-film level. The fitted atom-resolved polarizabilities should therefore be viewed as a consistent extension of the original Thole/Jensen model that preserves its good macroscopic performance while providing a more rigorous molecular-level description.

\subsection{Graph Neural Networks}

To make the first-principles parameterization scalable, we trained a graph neural network to predict the fitted atom-resolved Thole/Jensen parameters $\alpha_p$ directly from molecular structure. Machine-learning surrogates are particularly useful in this setting, where DFT reference data are accurate but expensive and therefore limited in size. Related surrogate and structure-dependent electrostatic models have been
introduced for distributed charge representations of molecular electrostatics~\cite{boittier2024kernel,boittier2022molecular,devereux2020polarizable}. In contrast, the present model learns local Thole/Jensen polarization parameters, i.e., atom-resolved polarizabilities,
rather than distributed charges or electrostatic potentials. The network does not predict molecular polarizabilities directly. Instead, it approximates the DFT-based fitting step by providing local site parameters that are subsequently used in the Thole/Jensen induced-dipole model.

The architecture follows an EGNN-inspired message-passing scheme~\cite{satorras2021n} combined with local reference frames~\cite{lippmann2024beyond,filling2025direct}. Molecules are represented as cutoff graphs with scalar atom embeddings $\bm{h}_i^{(0)}$ and radial basis features $\bm{\rho}(r_{ij})$. For each atom $i$, a local frame $\bm{R}_i$ is constructed from the neighboring geometry. Edge directions are then expressed in the receiver frame,
\begin{equation}
    \bm{u}_{ij}^{(i)}
    =
    \bm{R}_i^\mathrm{T}
    \dfrac{\bm{r}_i-\bm{r}_j}{|\bm{r}_i-\bm{r}_j|}.
\end{equation}
The scalar model therefore contains directional information, which is important for polarizability because the local electronic response depends not only on atom type and distance, but also on the orientation of the chemical environment.

For message passing, we define the scalar interatomic distance $r_{ij}=|\bm{r}_i-\bm{r}_j|$. Together with the radial basis features $\bm{\rho}(r_{ij})$ and the local-frame direction $\bm{u}_{ij}^{(i)}$, it forms the geometric edge descriptor
\begin{equation}
    \bm{e}_{ij}
    =
    \left[
    r_{ij},
    \bm{\rho}(r_{ij}),
    \bm{u}_{ij}^{(i)}
    \right].
\end{equation}
Here, $[\cdot]$ denotes concatenation of input features. The descriptor $\bm{e}_{ij}$ is therefore a feature vector containing radial and directional information for the edge $j\rightarrow i$.

At layer $l$, a message from atom $j$ to atom $i$ is computed by a learned edge network $\phi_e$,
\begin{equation}
    \bm{m}_{ij}^{(l)}
    =
    \phi_e
    \left(
    \bm{h}_i^{(l)},
    \bm{h}_j^{(l)},
    \bm{e}_{ij}
    \right)
    \odot
    \bm{g}_{ij}^{(l)} .
\end{equation}
Here, $\bm{h}_i^{(l)}$ and $\bm{h}_j^{(l)}$ are the scalar node features of the receiving and sending atoms, respectively, and $\bm{g}_{ij}^{(l)}$ is a learned gate that modulates the edge message element-wise. Messages are summed over the neighborhood $\mathcal{N}(i)$ of atom $i$,
\begin{equation}
    \bm{m}_{i}^{(l)}
    =
    \sum_{j\in\mathcal{N}(i)}
    \bm{m}_{ij}^{(l)} ,
\end{equation}
where $\mathcal{N}(i)$ contains all atoms connected to atom $i$ within the cutoff graph. The node features are then updated with a learned node network $\phi_h$ and a residual connection,
\begin{equation}
    \bm{h}_{i}^{(l+1)}
    =
    \bm{h}_{i}^{(l)}
    +
    \phi_h
    \left(
    \bm{h}_{i}^{(l)},
    \bm{m}_{i}^{(l)}
    \right).
\end{equation}
A final node-level readout maps the updated atom features to one scalar atomic polarizability per atom. Here, $\phi_e$ and $\phi_h$ denote trainable neural networks corresponding to the edge-message and node-update functions, respectively. In contrast to a standard EGNN, atomic coordinates are not updated. The optimized molecular geometry is fixed, and directional information enters through the local-frame features.

\begin{figure}[t]
    \centering
    \includegraphics[width=\columnwidth]{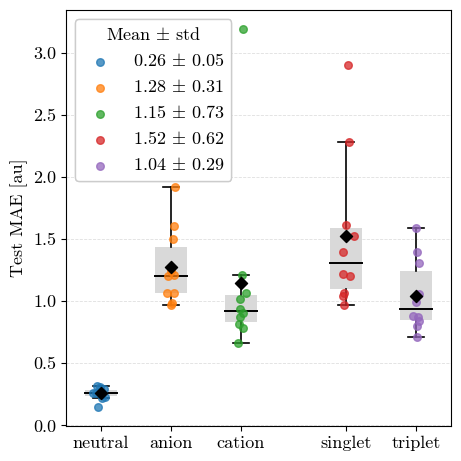}
    \caption{
    Atom-wise test MAE of the GNN-predicted atom-resolved polarizabilities
    for neutral, charged, and excited-state datasets. Each point corresponds
    to one independent train/validation/test split; black markers indicate
    the mean and standard deviation over all runs.
    }
    \label{fig:mae_average}
\end{figure}
The model was trained on DFT-derived atom-resolved polarizabilities using an 80/10/10 train/val/test split. As shown in Figure~\ref{fig:mae_average}, the prediction error is lowest for neutral molecules and increases for charged and excited states. Additional atom-wise parity plots for individual runs are shown in the Supporting Information.
This behavior is physically meaningful. Neutral ground-state molecules provide a comparatively smooth mapping between local atomic environment and fitted polarizability. The electronic density is well defined, charge redistribution is moderate, and chemically similar environments lead to similar local response parameters. In contrast, cations and anions depend sensitively on where the removed or added charge is localized. Excited states are even more state-specific, because diffuse densities, charge-transfer character, and spin-dependent localization can strongly change the local response.

Importantly, the neutral-state error of $0.26\pm0.05~\mathrm{au}$ is obtained from a very small first-principles data set, showing that the local-frame GNN captures the dominant structure--response relation efficiently. The larger errors for charged and excited states should therefore not be viewed primarily as a failure of the model, but as evidence that these states encode more non-local and state-dependent electronic response. Thus, the GNN does not yet fully replace the direct DFT-based fit, but it provides a promising and data-efficient surrogate for fitted atom-resolved Thole/Jensen parameters.

Unstable or unphysical reference cases were removed from the ML analysis before training and evaluation. A molecule was excluded if the DFT molecular polarizability exceeded $100\,000~\mathrm{au}$, if the deviation in the mean molecular polarizability was larger than $4\,000~\mathrm{au}$, or if at least one fitted atomic polarizability fell outside the range $|\alpha_p|\leq 50~\mathrm{au}$. These criteria identify cases where either the electronic-structure reference or the subsequent atom-resolved fit becomes ill-conditioned rather than physically meaningful. This is particularly relevant for excited states, where charge-transfer character, triplet instabilities, or strongly delocalized orbitals can lead to unrealistically large response values.

The filtering affected only a small subset of systems, but was essential to prevent the network from learning numerical artifacts instead of transferable local response patterns. The excluded cation was mCBP. The excluded singlet systems were CzDBA, mCBP, BCPO, TBA, TPTPA, TBPT, TCP, and 2-TNATA. The excluded triplet systems were TCTA, mCBP, TBA, TPTPA, TBPT, TCP, and 2-TNATA. Overall, the GNN provides an efficient and interpretable surrogate for atom-resolved Thole/Jensen parameterization. Its strong performance for neutral molecules is particularly notable given the small number of DFT reference systems, while the larger errors for charged and excited states reflect the increasingly non-local and state-dependent nature of the underlying electronic response.

\section{Conclusions and Outlook}

A central message of this work is that \textit{charge flow} (the field-induced redistribution of partial charges) is not a small correction but an essential component of the molecular response that must be treated explicitly when fitting \textit{atom-resolved} polarizabilities. In a distributed multipole picture, the external field induces not only atomic dipoles but also changes in atomic charges, giving rise to an additional \textit{charge-flow} contribution to the total induced dipole. While classical Thole/Jensen parameterizations can capture this effect \textit{implicitly} through fits to molecular polarizabilities, such an implicit treatment breaks down once the number of fitted parameters is increased to atom-specific (and potentially tensorial) values. We therefore incorporate charge flow directly by converting the induced charge response into atomic charge-flow dipoles and including them alongside the induced dipoles in the fitting functional.

Building on this physically complete fitting target, we introduced a first-principles parameterization strategy that replaces transferable, atom-type polarizabilities by \textit{molecule-specific, atom-resolved} parameters and, where needed, extends the description from scalar to tensorial atomic response. This increased resolution improves the representation of both the magnitude and the directional character of polarization, while leaving the computational cost of the induced-dipole solve unchanged once parameters are available.

Benchmarking against DFT-derived molecular polarizability tensors demonstrates that molecule-specific atomic polarizabilities substantially enhance agreement across chemically diverse systems. Neutral molecules are reproduced near-parity over most of the dataset, with the remaining deviations concentrated in very small species where restricting the local response to scalar sites becomes intrinsically underdetermined. Extending the atomic response to tensors resolves this bottleneck and systematically improves the anisotropy description, highlighting that directionality is not a simple global rescaling problem but a local, environment-dependent property.

A key outcome of this study is that the same parameterization framework remains applicable beyond neutral ground states. Charged states pose a more stringent test of transferability because polarization couples more strongly to both the local chemical environment and long-range induction. Here, atom-type parameterizations degrade markedly, whereas atom-resolved parameters largely restore predictive accuracy for both cations and anions. In our dataset, reproducing anisotropy remains slightly more challenging for anions, consistent with a stronger environment dependence of their electronic response, but the extended anisotropic model significantly narrows this gap.

We further showed that electronically excited states require an adaptive, state-sensitive description of atomic response. In excited manifolds, diffuse densities and charge-transfer character can render atom-type assignments untenable; incorporating molecule-specific polarizabilities and explicitly accounting for charge redistribution (charge-flow contribution) becomes essential to avoid systematic errors in molecular polarizability predictions. Taken together, these results establish atom-resolved (and, when necessary, tensorial) polarizabilities as a practical route to extend Thole/Jensen-type polarization models to charged and excited-state applications within a unified framework.

The conformer analysis of TPBi further clarifies the transferability of fitted atomic polarizabilities. Parameters fitted at a single reference geometry remain highly transferable for the mean molecular polarizability, especially in the tensorial model, whereas the molecular anisotropy is more sensitive to conformational changes and is best described by the atom-resolved scalar fit for the selected conformer set. This shows that conformer tests provide a useful diagnostic for distinguishing response properties that can be captured by a single set of fitted atomic polarizabilities from those that may require conformation-dependent or machine-learned parameters.

Beyond molecular benchmarks, we updated the refractive-index anisotropy workflow for simulated vapor-deposited thin films by replacing the previously used atom-type polarizabilities with the fitted atomic polarizabilities obtained here. This preserves the functional form of the polarization model while ensuring that the optical response is consistent with the electronic-structure level employed throughout this study. The atomistic morphologies entering these calculations are generated using the coarse-grained deposition and backmapping workflow introduced in our earlier work, providing a multiscale connection between molecular response parameters and film-scale optical observables.

Looking forward, two directions appear particularly promising. First, extending the reference data and fit targets to include additional conformers, chemical motifs, and higher charge states will further clarify the limits of scalar versus tensorial descriptions and enable more robust transferability. Second, machine-learning surrogates (e.g., equivariant GNNs) trained on first-principles reference data offer a scalable path to predict atom-resolved polarizabilities directly from structure, enabling high-throughput parameterization without compromising the physical interpretability of the underlying polarization model. Together, these developments pave the way toward predictive, polarizable force fields that remain accurate across heterogeneous environments and electronic states, thereby supporting simulation-driven design of organic electronic materials.


\section{Methods}
\subsection{Density functional calculations}
For the calculations of the ground state of neutral and simple charged molecules the induced dipoles were obtained using the software GAUSSIAN16 \cite{g16} with DFT-methods at the M06-2X/6-311+g(d,p) level of theory with and without an external applied field. To guarantee a reasonable comparison the same level of theory is used within the TDDFT-methods for the excited state calculations \cite{liang_2022}. The first singlet (S$_1$) and first triplet (T$_1$) excited states were selected as representative cases due to their relevance to optical absorption and emission processes in organic electronic materials. To suppress contributions from the dipole hyperpolarizability, which can be significant for diffuse excited state densities, a reduced static electric field strength of $E = 0.0001\ \mathrm{au}$ was employed. This value was verified to lie within the linear response regime and to yield field-independent polarizabilities. The same field protocol (magnitude and directions along $x$, $y$, and $z$ axes) was applied consistently across all electronic states to ensure direct comparability.

\subsection{Distributed multipole analysis}
The charge densities were partitioned using DDEC6 (Chargemol version $09\_26\_2017$) and the resulting charges and dipoles were subtracted to yield the induced quantities $\bm{\mu}_p^{\text{qm}}$ and $q_p^{\text{qm}}$ as explained in Section \textit{The fitting and parameterization algorithm} and used in Eq.~\ref{fit-function_final}.

\section{Data Availability}
The data sets that support the simulations reported here will be made freely accessible upon publication. The data set, complete workflow and all accompanying scripts of the atomic polarisabilities fitting are provided documented in the Git repository of SPARK (Simulation of Polarizabilities via Atomic Response Kernels) at XXX. 
The graph neural network including the training data is available at XXX.


\section{Acknowledgments}
This work was supported by the Deutsche Forschungsgemeinschaft (DFG, German Research Foundation) through the framework of the collaborative research center TRR 146 (Project No. 233630050). 

\bibliography{literature}

\end{document}